\documentclass[pre,aps,preprint,showpacs,preprintnumbers,amsmath,amssymb,secnumroman,eqsecnum,eqappnum]{revtex4}

\usepackage{graphicx}
\usepackage{dcolumn}
\usepackage{bm}

\newcommand{\beq}{\begin{equation}}
\newcommand{\eeq}{\end{equation}}
\newcommand{\beqa}{\begin{eqnarray}}
\newcommand{\eeqa}{\end{eqnarray}}
\newcommand{\vc}[1]{\mbox{\boldmath $#1$}}

\newcommand{\vol}[1]{{\bf #1}}

\newcommand{\du}[1]{{\bf\sf #1}}

\begin{document}


\title{Swimming of a sphere in a viscous incompressible fluid with inertia}

\author{B. U. Felderhof}

 \email{ufelder@physik.rwth-aachen.de}
\affiliation{Institut f\"ur Theorie der Statistischen Physik\\ RWTH Aachen University\\
Templergraben 55\\52056 Aachen\\ Germany\\
}%

\author{R. B. Jones}

 \email{r.b.jones@qmul.ac.uk}
\affiliation{Queen Mary University of London, The School of
Physics and Astronomy, Mile End Road, London E1 4NS, UK\\}%

\date{\today}

\begin{abstract}
The swimming of a sphere immersed in a viscous incompressible fluid with inertia is studied for surface modulations of small amplitude on the basis of the Navier-Stokes equations. The mean swimming velocity and the mean rate of dissipation are expressed as quadratic forms in term of the surface displacements. With a choice of a basis set of modes the quadratic forms correspond to two Hermitian matrices. Optimization of the mean swimming velocity for given rate of dissipation requires the solution of a generalized eigenvalue problem involving the two matrices. It is found for surface modulations of low multipole order that the optimal swimming efficiency depends in intricate fashion on a dimensionless scale number involving the radius of the sphere, the period of the cycle, and the kinematic viscosity of the fluid.
\end{abstract}

\maketitle
\section{\label{I}Introduction}

The swimming of fish and the flying of birds still pose problems to theory \cite{1}. The analysis can be based on the Navier-Stokes equations for the flow of a viscous fluid with a no-slip boundary condition at the surface of the body with periodically changing shape. For simplicity the fluid may be taken to be incompressible. The fluid is then characterized by its shear viscosity and mass density.

Most of the theoretical work has been concerned with either of two limiting situations. The swimming of microorganisms is well described by the time-independent Stokes equations of low Reynolds number hydrodynamics \cite{2}\cite{3}. The work in this area has been reviewed by Lauga and Powers \cite{4}. In the opposite limit of inviscid flow the analysis is based on the Euler equations with the effect of viscosity relegated to a boundary layer. The flow is predominantly irrotational, apart from the boundary layer and a wake of vorticity. The work in this field was reviewed by Sparenberg \cite{5}\cite{6} and by Wu \cite{7}\cite{8}. The modeling of bird flight was reviewed by Pennycuick \cite{9} and by Shyy et al. \cite{10}. The problem has also been addressed in computer simulation \cite{11}.

It is important to have a model covering the full range of kinematic viscosity. The seminal work of Taylor \cite{12} on the swimming of a sheet in the Stokes limit was extended to a fluid with inertia by Reynolds \cite{13} and by Tuck \cite{14}. The swimming of a planar slab has also been studied in the full range \cite{15}. The disadvantage of these models is the infinite length of the system which precludes study of the finite size effects which are believed to be crucial in the inviscid limit.

In earlier work we have studied small amplitude swimming of a body in a viscous fluid with inertia from a general point of view \cite{16}. As an example we studied the swimming of a sphere by means of potential flow \cite{17}. Later we showed that in the Stokes limit the addition of viscous modes leads to a significantly enhanced optimal efficiency \cite{18}. In the following we study the swimming of a sphere in the full range of kinematic viscosity. For simplicity we assume axisymmetric flow.

The effect of Reynolds stress turns out to be quite important. At small values of the kinematic viscosity it largely cancels the effect of viscous stress and pressure in the optimal efficiency. As a consequence the efficiency varies little with kinematic viscosity in a wide range. Similar behavior was found for an assembly of interacting rigid spheres \cite{19} and for a planar slab \cite{15}. We find that in swimming of a single sphere with surface distortions consisting of a running wave of three or five low order multipolar modes the optimal efficiency in the inviscid limit tends to the value for potential swimming. We expect that this feature holds more generally.

\section{\label{II}Flow equations}

We consider a flexible sphere of radius $a$ immersed in a viscous
incompressible fluid of shear viscosity $\eta$ and mass density $\rho$. In the laboratory frame, where the fluid is at rest at infinity, the flow velocity
$\vc{v}(\vc{r},t)$ and the pressure $p(\vc{r},t)$ satisfy the
Navier-Stokes equations
\begin{equation}
\label{2.1}\rho\bigg(\frac{\partial\vc{v}}{\partial t}+(\vc{v}\cdot\nabla)\vc{v}\bigg)=\eta\nabla^2\vc{v}-\nabla p,\qquad\nabla\cdot\vc{v}=0.
\end{equation}
The fluid is set in motion by time-dependent distortions of the
sphere. We shall study axisymmetric periodic distortions which lead to a translational swimming
motion of the sphere in the $z$ direction in a Cartesian system of coordinates. The surface displacement
$\vc{\xi}(\vc{s},t)$ is defined as the vector distance
\begin{equation}
\label{2.2}\vc{\xi}=\vc{s}'-\vc{s}
\end{equation}
of a point $\vc{s}'$ on the displaced surface $S(t)$ from the
point $\vc{s}$ on the sphere with surface $S_0$. The fluid
velocity $\vc{v}(\vc{r},t)$ in the rest frame is required to satisfy
\begin{equation}
\label{2.3}\vc{v}(\vc{s}+\vc{\xi}(\vc{s},t))=\frac{\partial\vc{\xi}(\vc{s},t)}{\partial t}.
\end{equation}
This amounts to a no-slip boundary condition. The instantaneous translational swimming velocity $\vc{U}(t)$ and the flow pattern $(\vc{v},p)$ follow from the condition that no net
force is exerted on the fluid.

We perform a perturbation expansion in powers of the displacement $\vc{\xi}(\vc{s},t)$.
The first order flow velocity $\vc{v}_1$ and pressure $p_1$ satisfy the linearized Navier-Stokes equations. The translational swimming velocity of the sphere, averaged over a period, is denoted by $\overline{\vc{U}}$. To first order in displacements the mean swimming velocity $\overline{\vc{U}}_1$ vanishes. We have shown previously \cite{16} that to second order the mean swimming velocity may be calculated as the sum of a surface and a bulk contribution
\begin{equation}
\label{2.4}
\overline{\vc{U}}_2=\overline{\vc{U}}_{2S}+\overline{\vc{U}}_{2B}.
\end{equation}
In spherical coordinates $(r,\theta,\varphi)$ the surface contribution $\overline{\vc{U}}_{2S}$ may be expressed as an integral of a mean surface velocity $\overline{\vc{u}}_S(\theta)$, defined by
\begin{equation}
\label{2.5}
\overline{\vc{u}}_S(\theta)=-\overline{(\vc{\xi}\cdot\nabla)\vc{v}_1}\big|_{r=a},
\end{equation}
where the overhead bar indicates the time average over a period. The surface contribution $\overline{\vc{U}}_{2S}$ to the swimming velocity is given by the spherical average \cite{17}
\begin{equation}
\label{2.6}
\overline{\vc{U}}_{2S}=-\frac{1}{4\pi}\int\overline{\vc{u}}_S(\theta)\;d\Omega.
\end{equation}
The bulk contribution $\overline{\vc{U}}_{2B}$ corresponds to the term $\rho(\vc{v}_1\cdot\nabla)\vc{v}_1$ in the Navier-Stokes equations. The time-averaged second order flow velocity $\overline{\vc{v}}_2$ and pressure $\overline{p}_2$ satisfy the inhomogeneous Stokes equations \cite{16}
\begin{equation}
\label{2.7}
\eta\nabla^2\overline{\vc{v}}_2-\nabla\overline{p}_2=\rho\overline{(\vc{v}_1\cdot\nabla)\vc{v}_1},\qquad\nabla\cdot\overline{\vc{v}_2}=0,
\end{equation}
with boundary condition
\begin{equation}
\label{2.8}
\overline{\vc{v}}_2\big|_{r=a}=\overline{\vc{u}}_S(\theta).
\end{equation}
The right hand side in Eq. (2.7) represents a force density $\overline{\vc{F}}_V=-\rho\overline{(\vc{v}_1\cdot\nabla)\vc{v}_1}$. The bulk part of the second order flow $\overline{\vc{v}}_{2B},\overline{p}_{2B}$ satisfies Eq. (2.7) with the no-slip boundary condition $\overline{\vc{v}}_{2B}\big|_{r=a}=0$. The contribution $\overline{\vc{U}}_{2B}$ to the mean swimming velocity follows from the flow velocity at infinity in the rest frame and the condition that no net force is exerted on the fluid. The integral of the force density $\overline{\vc{F}}_V$ is canceled by the surface integral of an induced force density on the sphere at rest. As we shall show later, the bulk contribution $\overline{\vc{U}}_{2B}$ may be calculated with the aid of an antenna theorem \cite{20}.

In the following we consider in particular the case of harmonic time variation. It is then convenient to introduce complex notation. The surface displacement is written as
\begin{equation}
\label{2.9}
\vc{\xi}(\theta,t)=\mathrm{Re}[\vc{\xi}_\omega(\theta)e^{-i\omega t}],
\end{equation}
with complex amplitude $\vc{\xi}_\omega(\theta)$. The corresponding first order flow velocity and pressure are given by
\begin{equation}
\label{2.10}
\vc{v}_1(\vc{r},t)=\mathrm{Re}[\vc{v}_\omega(\vc{r})e^{-i\omega t}],\qquad p_1(\vc{r},t)=\mathrm{Re}[p_\omega(\vc{r})e^{-i\omega t}].
\end{equation}
The time-averaged surface velocity $\overline{\vc{u}}_S(\theta)$ may be expressed as
\begin{equation}
\label{2.11}
\overline{\vc{u}}_S(\theta)=-\frac{1}{2}\mathrm{Re}[(\vc{\xi}^*_\omega\cdot\nabla)\vc{v}_\omega\big|_{r=a}].
\end{equation}
The time-averaged second order rate of energy dissipation is given by \cite{16}
\begin{equation}
\label{2.12}
\overline{\mathcal{D}}_2=-\frac{1}{2}\mathrm{Re}[\int_{S_0}\vc{v}^*_\omega\cdot\vc{\sigma}_\omega\cdot\vc{e}_r\;dS],
\end{equation}
where $\vc{\sigma}_\omega$ is the first order stress tensor with Cartesian components
\begin{equation}
\label{2.13}
\sigma_{\omega,\alpha\beta}=\eta\bigg(\frac{\partial v_{\omega\alpha}}{\partial x_\beta}+\frac{\partial v_{\omega\beta}}{\partial x_\alpha}\bigg)-p_\omega\delta_{\alpha\beta}.
\end{equation}

The efficiency of swimming is defined as
\begin{equation}
\label{2.14}
E_T(\omega)=4\eta\omega a^2\frac{|\overline{U}_2|}{\overline{\mathcal{D}}_2}.
\end{equation}
This quantity may be expressed as a ratio of two forms which are quadratic in the surface displacement, and involve two Hermitian matrices. Earlier we have studied the zero frequency limit of the problem where inertia plays no role \cite{18}. It has been shown by Shapere and Wilczek \cite{18A} that in this limit the above definition of efficiency is preferable to that of Lighthill \cite{18B}. The efficiency defined in Eq. (2.14) is essentially the ratio of speed and power and is relevant in the whole range of scale number.

\section{\label{III}Matrix formulation}

The explicit calculation requires the choice of a basis set of solutions of the linearized Navier-Stokes equations. After removal of the exponential time-dependent factor the equations read
\begin{equation}
\label{3.1}\eta[\nabla^2\vc{v}_\omega-\alpha^2\vc{v}_\omega]-\nabla p_\omega=0,\qquad\nabla\cdot\vc{v}_\omega=0,
\end{equation}
with the variable
\begin{equation}
\label{3.2}\alpha=(-i\omega\rho/\eta)^{1/2}=(1-i)(\omega\rho/2\eta)^{1/2}.
\end{equation}
In our previous work \cite{17} we have chosen a set of modes identical to those used earlier in a hydrodynamic scattering theory of flow about a sphere \cite{21}. In the present axisymmetric case we can use a reduced set of solutions. Moreover, it turns out that in the numerical work it is advantageous to use a different normalization. Thus we use the modes
 \begin{eqnarray}
\label{3.3}\vc{v}_l(\vc{r},\alpha)&=&\frac{2}{\pi}\;e^{\alpha a}[(l+1)k_{l-1}(\alpha r)\vc{A}_l(\hat{\vc{r}})+lk_{l+1}(\alpha r)\vc{B}_l(\hat{\vc{r}})],\nonumber\\
\vc{u}_l(\vc{r})&=&-\bigg(\frac{a}{r}\bigg)^{l+2}\vc{B}_l(\hat{\vc{r}}),\qquad p_l(\vc{r},\alpha)=\eta\alpha^2a\bigg(\frac{a}{r}\bigg)^{l+1}P_l(\cos\theta),
\end{eqnarray}
with modified spherical Bessel functions \cite{22} $k_l(z)$ and vector spherical harmonics $\{\vc{A}_l,\vc{B}_l\}$ defined by
 \begin{eqnarray}
\label{3.4}\vc{A}_l&=&\hat{\vc{A}}_{l0}=lP_l(\cos\theta)\vc{e}_r-P^1_l(\cos\theta)\vc{e}_\theta,\nonumber\\
\vc{B}_l&=&\hat{\vc{B}}_{l0}=-(l+1)P_l(\cos\theta)\vc{e}_r-P^1_l(\cos\theta)\vc{e}_\theta,
\end{eqnarray}
with Legendre polynomials $P_l$ and associated Legendre functions $P^1_l$ in the notation of Edmonds \cite{23}. The notation $\hat{\vc{A}}_{l0},\hat{\vc{B}}_{l0}$ is identical to that used in Ref. 26. In particular $\vc{A}_1=\vc{e}_z$ and $\vc{B}_1=\vc{e}_z-3\cos\theta\;\vc{e}_r$. The solutions $\vc{v}_l(\vc{r},\alpha)$ are associated with vanishing pressure variation. The notation for the flow field $\vc{u}_l(\vc{r})$ is identical to that in our previous work for zero frequency \cite{18}. At zero frequency there is no pressure variation associated with these irrotational flow fields. We remark here that the above basis set is not suitable at low frequency owing to the singularity of the solutions $\vc{v}_l(\vc{r},\alpha)$ at $\alpha=0$. At low frequency we must use a modified set, as discussed later.

We expand the first order flow velocity and pressure in the modes, given by Eq. (3.3), as
\begin{equation}
\label{3.5}\vc{v}_\omega(\vc{r})=-\omega a\sum^\infty_{l=1}[\kappa_l\vc{v}_l(\vc{r},\alpha)+\mu_l\vc{u}_l(\vc{r})],\qquad p_\omega(\vc{r})=-\omega a\sum^\infty_{l=1}\mu_lp_l(\vc{r},\alpha),
\end{equation}
with complex coefficients $\{\kappa_l,\mu_l\}$, which can be calculated as moments of the function $\vc{v}_\omega(\vc{r})$ on the surface $r=a$. Correspondingly the displacement function $\vc{\xi}_\omega(\hat{\vc{r}})$ is expanded as
\begin{equation}
\label{3.6}\vc{\xi}_\omega(\hat{\vc{r}})=-ia\sum^\infty_{l=1}[\kappa_l\vc{v}_l(\vc{s},\alpha)+\mu_l\vc{u}_l(\vc{s})].
\end{equation}
We define the complex multipole moment vector $\vc{\psi}$ as the one-dimensional array
\begin{equation}
\label{3.7}\vc{\psi}=(\kappa_1,\mu_1,\kappa_2,\mu_2,....).
\end{equation}
Then $\overline{U_2}$ can be expressed as
\begin{equation}
\label{3.8}\overline{U_2}=\frac{1}{2}\omega
a(\vc{\psi}|\du{B}|\vc{\psi}),
\end{equation}
with a dimensionless Hermitian matrix
$\du{B}$ and the notation
\begin{equation}
\label{3.9}(\vc{\psi}|\du{B}|\vc{\psi})=\sum_{l,\sigma,l',\sigma'}\psi^*_{l\sigma}B_{l\sigma,l'\sigma'}\psi_{l'\sigma'},
\end{equation}
where the subscript $\sigma$ takes the two values $N,P$ with $\psi_{lN}=\kappa_l$ and $\psi_{lP}=\mu_l$. The subscripts $N,P$ correspond to notation used in earlier work \cite{17},\cite{21} for modes proportional to those in Eq. (3.3). We impose the constraint that the force exerted on the fluid vanishes at any time. This requires $\kappa_1=0$. We implement the constraint by dropping the first
element of $\vc{\psi}$ and erasing the first row and column of the
matrix $\du{B}$. We denote the corresponding modified vector as
$\hat{\vc{\psi}}$ and the modified matrix as $\hat{\du{B}}$.

The time-averaged rate of dissipation can be expressed as
\begin{equation}
\label{3.10}\overline{\mathcal{D}_2}=8\pi\eta\omega^2a^3(\vc{\psi}|\du{A}|\vc{\psi}),
\end{equation}
with a dimensionless Hermitian matrix $\du{A}$. We denote
the modified matrix obtained by dropping the first row and column
by $\hat{\du{A}}$.

With the constraint $\kappa_1=0$ the mean swimming velocity
$\overline{U}_2$ and the mean rate of dissipation $\overline{\mathcal{D}}_2$ can be expressed as
\begin{equation}
\label{3.11}\overline{U_2}=\frac{1}{2}\omega
a(\hat{\vc{\psi}}|\hat{\du{B}}|\hat{\vc{\psi}}),\qquad\overline{\mathcal{D}_2}=8\pi\eta\omega^2a^3
(\hat{\vc{\psi}}|\hat{\du{A}}|\hat{\vc{\psi}}).
\end{equation}
Optimization of the mean swimming velocity for given mean rate of dissipation, taken into account with a Lagrange multiplier $\lambda$, leads to the eigenvalue problem
\begin{equation}
\label{3.12}\hat{\du{B}}|\hat{\vc{\psi}}_\lambda)=\lambda\hat{\du{A}}|\hat{\vc{\psi}}_\lambda).
\end{equation}
Both matrices $\hat{\du{B}}$ and $\hat{\du{A}}$ are hermitian, so that the eigenvalues $\lambda$ are real. With truncation at maximum $l$-value $L$ the
truncated matrices $\hat{\du{A}}_{1L}$ and $\hat{\du{B}}_{1L}$ are $2L-1$-dimensional. The maximum positive eigenvalue $\lambda_{max}$ is of particular interest. Its corresponding eigenvector provides the swimming mode of maximal efficiency. The truncated matrices correspond to swimmers obeying the constraint that all multipole coefficients for $l>L$ vanish.

With use of Eq. (2.6) we find the contribution $\du{B}_S$ to the matrix $\du{B}=\du{B}_S+\du{B}_B$. The remainder $\du{B}_B$ follows from the contribution $\overline{\vc{U}}_{2B}$ to the mean swimming velocity. The elements of the matrices $\du{B}_S$ and $\du{A}$ are complex numbers which can be calculated by substitution of the expansions in Eq. (3.5) and (3.6) into the expressions (2.11) and (2.12). We have calculated the elements of the matrices $\du{B}_S$ and $\du{A}$ in our earlier work \cite{17}. In order to make contact with our subsequent work on the axisymmetric case in the limit of zero frequency \cite{18} we have performed an independent calculation for axial symmetry. In the calculation we use angular integrals which are detailed in Appendix A.

The matrix $\du{A}$ is diagonal in $l,l'$, and the matrices $\du{B}_S$ and $\du{B}_B$ are tridiagonal in $l,l'$. The matrices are frequency-dependent via the variable $\alpha a$. We write
\begin{equation}
\label{3.13}\alpha a=(1-i)s,\qquad s=a\sqrt{\frac{\omega\rho}{2\eta}}=\frac{1}{\sqrt{2\eta_*}},\qquad\eta_*=\frac{\eta}{\omega\rho a^2},
\end{equation}
where $\eta_*$ is the dimensionless viscosity \cite{25}. The maximum eigenvalue $\lambda_{max}$ depends on the variable $s$. We call $s$ the scale number. It is related to the Roshko number $Ro=L^2f\rho/\eta$, where $f=\omega/(2\pi)$, by $Ro=4s^2/\pi$, if in $Ro$ we use the sphere diameter $2a$ as the characteristic length $L$.

From Eq. (3.13) we see that for given fluid properties the frequency must decrease with increasing radius as $1/a^2$ in order to keep the scale number $s$ constant. For water the kinematic viscosity takes the value $\eta/\rho=0.01\;\mathrm{cm}^2/\mathrm{sec}$, and in air it takes the value $\eta/\rho=0.15\;\mathrm{cm}^2/\mathrm{sec}$. Hence for air we have $s\approx5a\sqrt{f}$ with $a$ in cm and $f=\omega/(2\pi)$ in Hz.

\section{\label{IV}Effect of Reynolds stress}

The force density $\overline{\vc{F}}_V=-\rho\overline{(\vc{v}_1\cdot\nabla)\vc{v}_1}$ in Eq. (2.7) can be written alternatively as the divergence of a Reynolds stress $\overline{\vc{F}}_V=-\rho\nabla\cdot\overline{(\vc{v}_1\vc{v}_1)}$. In order to find the contribution to the mean swimming velocity caused by this stress we must solve Eq. (2.7) with no-slip boundary condition at $r=a$. We explained the procedure below Eq. (2.8). The mean swimming velocity $\overline{\vc{U}}_{2B}$ corresponds to the matrix $\du{B}_B$ as in Eq. (3.8).

According to theory developed earlier \cite{16} the second order swimming velocity corresponding to the mean Reynolds stress can be written as a sum of two contributions,
\begin{equation}
\label{4.1}\overline{\vc{U}}_{2B}=\overline{\vc{U}}_V+\widetilde{\overline{\vc{U}}}_V,
\end{equation}
where $\overline{\vc{U}}_V$ corresponds to the integral of the force density
\begin{equation}
\label{4.2}\vc{\mathcal{F}}_V=\int\overline{\vc{F}}_V\;d\vc{r},
\end{equation}
according to Stokes' law
\begin{equation}
\label{4.3}\overline{\vc{U}}_V=\frac{-1}{6\pi\eta a}\vc{\mathcal{F}}_V.
\end{equation}
The remainder $\widetilde{\overline{\vc{U}}}_V$ corresponds to a solution of the Stokes problem Eq. (2.7) corresponding to the force density $\overline{\vc{F}}_V$ and a freely moving sphere of radius $a$.

In order to find the contribution $\widetilde{\overline{\vc{U}}}_V$ we use an antenna theorem derived earlier \cite{20}. The force density $\overline{\vc{F}}_V$ is decomposed into vector spherical harmonics as
\begin{equation}
\label{4.4}\overline{\vc{F}}_V(\vc{r})=f_A(r)\vc{A}_1+f_B(r)\vc{B}_1+\delta\overline{\vc{F}}_V(\vc{r}),
\end{equation}
with scalar functions $f_A(r),f_B(r)$ given by the angular integrals
\begin{eqnarray}
\label{4.5}f_A(r)&=&\frac{1}{2}\int^\pi_0\vc{e}_z\cdot\overline{\vc{F}}_V(\vc{r})\sin\theta\;d\theta,\nonumber\\
f_B(r)&=&\frac{1}{4}\int^\pi_0(\vc{e}_z-3\cos\theta\;\vc{e}_r)\cdot\overline{\vc{F}}_V(\vc{r})\sin\theta\;d\theta,
\end{eqnarray}
and with remainder $\delta\overline{\vc{F}}_V(\vc{r})$ given by a sum of higher order vector spherical harmonics $\{\vc{A}_l,\vc{B}_l\}$.
According to the antenna theorem \cite{20} the Green function solution of the Stokes equations, assumed to be valid in all space in the absence of the sphere, corresponds to a flow velocity for $r<a$ given by
\begin{equation}
\label{4.6}\vc{v}_0(\vc{r})=c^+_{100}\vc{v}^+_{100}(\vc{r})+c^+_{102}\vc{v}^+_{102}(\vc{r})+\delta\vc{v}_0(\vc{r}),
\end{equation}
with flow fields $\vc{v}^+_{100}(\vc{r}),\;\vc{v}^+_{102}(\vc{r})$ given by \cite{24}
\begin{equation}
\label{4.7}\vc{v}^+_{100}(\vc{r})=\vc{A}_1=\vc{e}_z,\qquad\vc{v}^+_{102}(\vc{r})=r^2(5\vc{A}_1+\vc{B}_1),
\end{equation}
coefficients
\begin{equation}
\label{4.8}c^+_{100}=\frac{1}{3\eta}\int^\infty_a[2r'f_A(r')-r'f_B(r')]\;dr',\qquad c^+_{102}=\frac{1}{15\eta}\int^\infty_a\frac{1}{r'}f_B(r')\;dr',
\end{equation}
and remainder $\delta\vc{v}_0(\vc{r})$ given by a sum of higher order vector spherical harmonics $\{\vc{A}_l,\vc{B}_l\}$.
The Green function tends to zero at infinity.

The velocity $\widetilde{\overline{\vc{U}}}_V$ in Eq. (4.1) follows from Fax\'en's theorem as \cite{24}
\begin{equation}
\label{4.9}\widetilde{\overline{\vc{U}}}_V=(c^+_{100}+5a^2c^+_{102})\vc{e}_z.
\end{equation}
By construction the flow pattern $\overline{\vc{v}}_{2B}(\vc{r})$ tends to $-\overline{\vc{U}}_{2B}$ plus a flow which decays faster than $1/r$ at large distance from the origin.

\section{\label{V}Expansion at low frequency}

As we have noted in Sec. III the singular behavior of the flow fields $\vc{v}_l(\vc{r},\alpha)$ at $\alpha=0$ leads to numerical difficulties at low frequency \cite{21}. Thus instead of $\vc{v}_l(\vc{r},\alpha)$ we use
\begin{equation}
\label{5.1}\vc{v}^0_l(\vc{r},\alpha)=X_l(\alpha)\vc{v}_l(\vc{r},\alpha)+\frac{2(2l-1)}{\alpha^2a^2}\vc{u}_l(\vc{r}),
\end{equation}
with coefficient
\begin{equation}
\label{5.2}X_l(\alpha)=\frac{2\alpha^la^l}{l(2l+1)(2l-3)!!}\;e^{-\alpha a}.
\end{equation}
It may be checked that at zero frequency
\begin{equation}
\label{5.3}\vc{v}^0_l(\vc{r},0)=\vc{v}_l(\vc{r}),
\end{equation}
where $\vc{v}_l(\vc{r})$ is the viscous mode function
\begin{equation}
\label{5.4}
\vc{v}_l(\vc{r})=\bigg(\frac{a}{r}\bigg)^{l}\big[(l+1)P_l(\cos\theta)\vc{e}_r
+\frac{l-2}{l}P^1_l(\cos\theta)\vc{e}_\theta\big].
\end{equation}
used in the zero frequency theory \cite{18}.

The pressure corresponding to the velocity mode in Eq. (5.1) is given by
\begin{equation}
\label{5.5}p^0_l(\vc{r})=\frac{2(2l-1)}{\alpha^2a^2}p_l(\vc{r},\alpha).
\end{equation}
Correspondingly the expansion in Eq. (3.5) must be replaced by
 \begin{eqnarray}
\label{5.6}\vc{v}_\omega(\vc{r})&=&-\omega a\sum^\infty_{l=1}[\kappa^0_l\vc{v}^0_l(\vc{r},\alpha)+\mu^0_l\vc{u}_l(\vc{r})],\nonumber\\
p_\omega(\vc{r})&=&-\omega a\sum^\infty_{l=1}[\kappa^0_lp^0_l(\vc{r})+\mu^0_lp_l(\vc{r},\alpha)].
\end{eqnarray}

With multipole vector $\psi^0$ defined by
\begin{equation}
\label{5.7}\vc{\psi}^0=(\kappa^0_1,\mu^0_1,\kappa^0_2,\mu^0_2,....)
\end{equation}
the mean swimming velocity and mean rate of dissipation can be expressed as
\begin{equation}
\label{5.8}\overline{U_2}=\frac{1}{2}\omega
a(\vc{\psi}^0|\du{B}^0|\vc{\psi}^0),\qquad\overline{\mathcal{D}_2}=8\pi\eta\omega^2a^3\omega
a(\vc{\psi}^0|\du{A}^0|\vc{\psi}^0),
\end{equation}
with matrices $\du{B}^0$ and $\du{A}^0$.

The above expansion is not suitable at high frequency owing to numerical difficulties in the eigenvalue problem analogous to Eq. (3.13). Thus we must use two different expansions in the two regimes of low and high frequency. The eigenvalues for the two eigenvalue problems are of course the same, and the eigenvectors are related by the matrix transforming one basis set into the other. At zero frequency the two matrices in the representation of this section are identical to those derived earlier \cite{18}.

\section{\label{VI}Results}

With a small number of modes at low multipole order the calculations can be performed analytically. With higher multipoles included the generalized eigenvalue problem can be solved numerically by use of the Eigensystem command of Mathematica. In our earlier work \cite{17} we have performed calculations involving just potential modes. In that case the Reynolds stress vanishes. As we have shown at zero frequency \cite{18} the inclusion of viscous modes leads to a significantly higher maximum efficiency.

The qualitative behavior can be seen from a simple model with only modes of orders $l=1$ and $l=2$ included. In this case there are just three modes, the dipolar potential mode at $l=1$, the viscous mode at $l=2$, and the quadrupolar potential mode at $l=2$. The modes are given by Eq. (3.3) at high frequency, and by Eqs. (5.1) and (5.5) for the viscous modes at low frequency. It turns out that in this case the high frequency expansion works well numerically over the whole range of interest. It suffices to compare with the zero frequency results \cite{18}.

From Eqs. (7.11) and (7.17) of Ref. 18 the matrix $\hat{\du{B}}^0_{12}$ at zero frequency is given by
\begin{equation}
\label{6.1}\hat{\du{B}}^0_{12}=i\left(\begin{array}{ccc}0&-\frac{3}{5}&-3
\\\frac{3}{5}&0&0
\\3&0&0\end{array}\right),
\end{equation}
and the matrix $\hat{\du{A}}^0_{12}$ is given by
\begin{equation}
\label{6.2}\hat{\du{A}}^0_{12}=\left(\begin{array}{ccc}3&0&0
\\0&\frac{27}{10}&\frac{18}{5}
\\0&\frac{18}{5}&6\end{array}\right).
\end{equation}
We denote the corresponding maximum eigenvalue by $\lambda^0_{12}$. The eigenvalue problem yields \cite{26}
\begin{equation}
\label{6.3}\lambda^0_{12}=\frac{5}{3\sqrt{2}}\approx 1.17851.
\end{equation}
The corresponding eigenvector $\vc{\xi}^0_0$ has components $(1,-4i\sqrt{2}/3,11i/(5\sqrt{2})\approx(1,-1.886i,1.556i)$.

With fluid inertia included the maximum eigenvalue as a function of the scale number is given by the expression
\begin{equation}
\label{6.4}\lambda_{12}(s)=\bigg(\frac{N(s)}{D(s)}\bigg)^{1/2},
\end{equation}
with numerator
\begin{eqnarray}
\label{6.5}N(s)&=&225+450s+450s^2+282s^3-12s^4-24s^5+104s^6+16s^7+4s^8-8s^9+8s^{10}\nonumber\\
&+&16\mathrm{Re}\big[\big(6i-(6-6i)s-3s^2+(1+i)s^3-is^4-(1-i)s^5\big)s^6e^{s-is}E_1(s-is)\big]\nonumber\\
&+&16s^{12}e^{2s}|E_1(s-is)|^2,
\end{eqnarray}
and denominator
\begin{equation}
\label{6.6}D(s)=18(9+18s+18s^2+10s^3).
\end{equation}
The expression is derived by use of the characteristic equation from the matrices $\hat{\du{A}}_{12}(s)$ and $\hat{\du{B}}_{12}(s)$ given explicitly in Appendix B. The denominator is related to the determinant of $\hat{\du{A}}_{12}(s)$ by $D(s)=(8s^4/3)\mathrm{det}[\hat{\du{A}}_{12}(s)]$.

In Fig. 1 we plot the function $\lambda_{12}(s)$ as a function of $\log_{10}(s)$. The low frequency expansion is given by
\begin{equation}
\label{6.7}\lambda_{12}(s)=\frac{5}{3\sqrt{2}}+\frac{8\sqrt{2}}{135}s^3-\frac{19\sqrt{2}}{135}s^4
+\frac{16\sqrt{2}}{135}s^5+O(s^6),
\end{equation}
in accordance with Eq. (6.3). The function shows a maximum value $1.183$ at $s_0=0.865$, corresponding to the eigenvector $\vc{\xi}_0$ with components $(1,-0.218+0.130i,7.911+1.001i)$ for the modes of Eq. (3.3). For large values of $s$ the numerical calculation can be performed by a method discussed in the next section. At high scale number the contribution of the Reynolds stress to the mean swimming velocity is quite important. If it were omitted, then the maximum eigenvalue would grow as $4\sqrt{s}/(3\sqrt{5})$. Actually, the maximum eigenvalue tends to the limiting value
\begin{equation}
\label{6.8}\lambda^{pot}_{12}=\frac{1}{\sqrt{2}}\approx 0.70711,
\end{equation}
the optimal value for potential swimming with just the dipolar and quadrupolar modes. The eigenvalue $\lambda^{pot}_{12}$ and the corresponding eigenvector $\vc{\xi}_0=(1,0,i/\sqrt{2})$ are found from Eqs. (6.1) and (6.2) with the second row and column of the matrices deleted.

The maximum eigenvalue at given $s$ increases as more modes are included. In Fig. 2 we show the maximum eigenvalue $\lambda_{13}(s)$ as a function of $\log_{10}(s)$ for surface displacements given by a superposition of the five modes for $l=1,2,3$. In this case the numerical calculation is more demanding than for $L=2$. We use the modes of Sec. V for $s<2$ and the modes of Eq. (3.3) for $s>2$. The maximum eigenvalue at zero frequency is $\lambda^0_{13}=1.514$ and the corresponding eigenvector has components $\vc{\xi}^0_0=(1,-1.553i,1.824i,1.373,-1.440)$. There is a weak maximum $\lambda_{13}(s_0)=1.516$ at $s_0=0.962$. The corresponding eigenvector has components $\vc{\xi}_0=(1,-0.715+1.592i,0.262-1.861i,1.291+0.198i,-1.385-0.060i)$ in terms of the modes of Sec. V. As $s$ increases beyond $s_0$ the function $\lambda_{13}(s)$ decays to $\lambda^{pot}_{13}=\sqrt{11/10}\approx 1.049$, the optimal value for potential swimming with modes $l=1,2,3$ with corresponding eigenvector $\vc{\xi}_0=(1,0,\sqrt{11/10}i,0,-3/5)$.

\section{\label{VII}Behavior for large scale number}

The behavior shown in Figs. 1 and 2 requires further discussion of the asymptotic properties for large scale number. The function $\lambda_{13}(s)$ shown in Fig. 2 is calculated from a complicated analytic expression involving exponential integrals, like the expression Eq. (6.4) for $\lambda_{12}(s)$. The expression is derived from the explicit expressions for the matrices $\hat{\du{A}}_{13},\;\hat{\du{B}}_{S13}$ and $\hat{\du{B}}_{B13}$ given in Appendix B, by use of the characteristic equation.

The exponential integrals appear multiplied by exponentials. It is therefore useful to define
\begin{equation}
\label{7.1}F(z)=e^zE_1(z)=\int^\infty_0\frac{e^{-u}}{z+u}\;du.
\end{equation}
In the expression for $\lambda_{13}(s)$ we need the value of $F(z)$ at $z=s\pm is$ and at $z=2s$ for large positive $s$. Clearly the function $F(z)$ is analytic in the complex $z$ plane apart from a branch cut along the negative real axis. The values of $F(z)$ which are needed in Fig. 2 can be found accurately by numerical integration in Eq. (7.1). The expression for $\lambda_{13}(s)$ involves powers of $z$ up to $z^{12}$ and down to $z^{-10}$, so that in numerical calculations for large $s$ we need integer programming.

To scrutinize the behavior for large values of $s$ it is useful to derive series expansions. By integration by parts we derive
\begin{equation}
\label{7.2}F(z)=F_n(z)+R_n(z),
\end{equation}
where $F_n(z)$ is given by a sum of $n+1$ terms,
\begin{equation}
\label{7.3}F_n(z)=\sum^n_{j=0}(-1)^j\frac{j!}{z^{j+1}},
\end{equation}
and the remainder $R_n(z)$ is given by
\begin{equation}
\label{7.4}R_n(z)=-(n+1)!\int^\infty_0\frac{e^{-u}}{(z+u)^{n+2}}\;du.
\end{equation}
The sum $F_n(z)$ corresponds to the sum of the first $n+1$ terms in the asymptotic expansion of the exponential integral \cite{22}.

By replacing $F(z)$ by the sum $F_n(z)$ for low values of $n$ in the expressions for $\lambda_{12}(s)$ and $\lambda_{13}(s)$ we can derive series expansions in inverse powers of $s$.
For $\lambda_{12}(s)$ we find
\begin{equation}
\label{7.5}\lambda_{12}(s)=\frac{1}{\sqrt{2}}+\frac{16 \sqrt{2}}{5s}-\frac{112\sqrt{2}}{5s^2}+\frac{20736\sqrt{2}}{125s^3}+O(s^{-4}).
\end{equation}
At $s=100$ the sum $\lambda^{(4)}_{12}(s)$ of the first four terms shown agrees with the exact value to four decimal places.
For $\lambda_{13}(s)$ we find by the same method
\begin{equation}
\label{7.6}\lambda_{13}(s)=\sqrt{\frac{11}{10}}+\frac{128349}{2695}\sqrt{\frac{2}{55}}\;\frac{1}{s}+O(s^{-2}).
\end{equation}
At $s=100$ the first two terms of the expansion in Eq. (7.6) yield a value accurate to one percent and at $s=10^4$ the value is accurate to five decimal places.

The eigenvector corresponding to the maximum eigenvalue $\lambda_{13}(s)$ can also be found by the same numerical method. For example, at $s=10^4$ we find the eigenvector $\vc{\xi}_0=(1,-3.219+3.221i,-1.048i,1.311+1.312i,-0.600)$ for the modes of Sec. III. This shows that for this scale number all five modes contribute significantly to the eigenvector. The contributions from $\hat{\du{B}}_{S13}$ and $\hat{\du{B}}_{B13}$ cancel to a large extent. We find the ratios $|\overline{U}_{2S}|/|\overline{U}_{2B}|=1.472$ and $|\overline{U}_{2S}|/|\overline{U}_{2S}+\overline{U}_{2B}|=3.112$. This shows again the importance of the Reynolds stress.

We note that for these modes of low order the whole calculation can be performed analytically, though at the expense of quite complicated expressions. It follows from the expressions in Appendix B that the elements of the matrix $\hat{\du{B}}_{S13}(s)$ remain finite in the limit $s\rightarrow\infty$ and apparently there is a cancellation from the matrix $\hat{\du{B}}_{B13}(s)$ leading to the limiting behavior shown in Figs. 1 and 2.

\section{\label{VIII}Simple model}

It is worthwhile to discuss the properties of the simple model introduced at the beginning of Sec. VI in some more detail. We denote the coefficients of the zero frequency modes of Ref. 18 as $(\mu^I_1,\kappa^I_2,\mu^I_2)$. These are the amplitudes of the dipolar mode, the stresslet, and the quadrupolar mode, respectively. By comparison of the amplitudes of vector spherical harmonics on the spherical surface $r=a$ we find that the coefficients are related to the coefficients $(\mu_1,\kappa_2,\mu_2)$ of the modes of Sec. III by the relations
\begin{eqnarray}
\label{8.1}\mu_1&=&\mu^I_1,\qquad\kappa_2=\frac{1}{5}\;\frac{z^2}{1+z}\;\kappa^I_2,\nonumber\\
\mu_2&=&\frac{6+6z+3z^2+z^3}{z^2+z^3}\;\kappa^I_2+\mu^I_2,\qquad z=(1-i)s.
\end{eqnarray}
These coefficients give the same surface displacement as the zero frequency modes for the set $(\mu^I_1,\kappa^I_2,\mu^I_2)$.
It follows from Eq. (8.1) that if one considers surface displacements characterized by a chosen set of coefficients $(\mu^I_1,\kappa^I_2,\mu^I_2)$, then the coefficient $\kappa_2$ of the boundary layer mode grows in absolute magnitude beyond all bounds as the scale number $s$ increases, whereas the amplitudes of the potential modes remain bounded. This causes an increase of dissipation as $s$ increases, owing to steep gradients in the boundary layer. Hence the efficiency decreases with increasing $s$. As an example we show in Fig. 3 the ratio
\begin{equation}
\label{8.2}\rho_{12}(s)=\frac{(\vc{\xi}(s)|\hat{\du{B}}_{12}(s)|\vc{\xi}(s))}{(\vc{\xi}(s)|\hat{\du{A}}_{12}(s)|\vc{\xi})(s)},
\end{equation}
where $\vc{\xi}(s)$ is calculated from the vector $\vc{\xi}^0_0$ given below Eq. (6.3) by use of Eq. (8.1). The ratio starts at the optimal value $\lambda_{12}(0)$ in the limit $s\rightarrow 0$, but eventually decreases to zero at large $s$. In Fig. 4 we show the surface deformation at four chosen times.

The simplest way of constructing a good mode of swimming for large $s$ is to avoid the boundary layer altogether and use surface displacements corresponding to the potential mode $\vc{\xi}^{pot}_0=(1,0,i/\sqrt{2})$ given below Eq. (6.8). In this case $\mu_1=1,\;\kappa_2=0,\;\mu_2=i/\sqrt{2}$, independent of $s$. This is a mode of high efficiency, only slightly less efficient than the optimal mode given by the maximum eigenvector of Eq. (3.12).

\section{\label{IX}Discussion}

The analysis shows that for given surface distortion the mean swimming velocity and mean power of a sphere depend in an intricate way on the dimensionless scale number $s$, defined in Eq. (3.13) and related to the Roshko number by $Ro=4s^2/\pi$. Optimization of the mean swimming velocity for given power at fixed $s$ leads to a generalized eigenvalue problem. The eigenvector with largest eigenvalue within a class of strokes characterizes the optimal stroke in that class. Explicit expressions for the two Hermitian matrices characterizing the eigenvalue problem are given in Appendix B. The expressions are the pinnacle of the present analysis.

The results are of particular interest for large values of the scale number. It turns out that in this range it is crucial to take the Reynolds stress into account. Apparently on time average over a period the effect of pressure gradients is largely canceled by the effect of Reynolds stress. Quantitatively the effect is measured by a comparison of the contributions $\overline{U}_{2B}$ and $\overline{U}_{2S}$ to the mean swimming velocity. A numerical example is given at the end of Sec. VII.

In the asymptotic range of very large scale number $s$ we find that the optimal efficiency tends to the value for potential flow for the class of axisymmetric strokes involving the three modes of order $l=1,2$, as well as for axisymmetric strokes involving the five modes of order $l=1,2,3$. The same behavior was found for the swimming of a deformable slab \cite{15}. It may be conjectured that for a sphere this behavior occurs also for more complicated strokes. It is shown in Figs. 1 and 2 that in the asymptotic range the swimming is somewhat less effective than in the Stokes limit. There is an appreciable change in the mode of optimal swimming as is seen by a comparison of the eigenvectors of displacements at the end of Secs. VI and VII.

For the strokes considered the optimal efficiency shows a maximum in the intermediate range of scale number. In this regime the optimal efficiency varies little with the scale number. For a swimmer the actual value of the scale number is determined by its geometrical dimension and the kinematic viscosity of the fluid. For that particular value of $s$ the swimmer can optimize its stroke to second order in the amplitude from the solution of the eigenvalue problem. It will be of interest to determine how the efficiency varies as the amplitude of stroke is increased. This requires numerical solution of the Navier-Stokes equations with no-slip boundary condition for the moving surface and is beyond the scope of the present investigation.
\appendix

\newpage

\section{\label{A}}
In this Appendix we provide expressions for angular integrals which occur in the calculation. We consider vector-functions of the form
  \begin{equation}
\label{A.1}\vc{v}_l(\vc{r})=f(r)\vc{A}_l+g(r)\vc{B}_l,
\end{equation}
and scalar functions of the form
  \begin{equation}
\label{A.2}p_l(\vc{r})=h(r)P_l(\cos\theta).
\end{equation}
In the calculation of the matrix $\du{A}$ we need the angular integral
  \begin{eqnarray}
\label{A.3}S_l(r)&=&\int^\pi_0\vc{v}^*_l(\vc{r})\cdot[\nabla\vc{v}_l(\vc{r})+\widetilde{\nabla\vc{v}}_l(\vc{r})]\cdot\vc{e}_r\sin\theta\;d\theta\nonumber\\
&=&\frac{2}{2l+1}\big[(l-1)l(l+1)\frac{(f^*+g^*)f}{r}-l(l+1)(l+2)\frac{(f^*+g^*)g}{r}\nonumber\\&-&l(l+1)(f^*g'+g^*f')+l(3l+1)f^*f'+(l+1)(3l+2)g^*g'\big],
\end{eqnarray}
where in the last equation we have omitted the dependence on $r$ for brevity. Similarly
  \begin{eqnarray}
\label{A.4}Q_l(r)&=&\int^\pi_0\vc{v}^*_l(\vc{r})p_l(r)\cdot\vc{e}_r\sin\theta\;d\theta\nonumber\\
&=&\frac{2}{2l+1}\big[lf^*h-(l+1)g^*h\big].
\end{eqnarray}
In the calculation of the matrix $\du{B}_S$ we need the angular integral
  \begin{eqnarray}
\label{A.5}T^{(1)}_l(r)&=&\int^\pi_0\vc{v}^*_l(\vc{r})\cdot\nabla\vc{v}_{l+1}(\vc{r})\cdot\vc{e}_z\sin\theta\;d\theta\nonumber\\
&=&\frac{2l+2}{2l+1}\big[l(l+1)\frac{(f^*+g^*)f}{r}+lf^*f'-(l+1)g^*f'\big],
\end{eqnarray}
as well as the integral
  \begin{eqnarray}
\label{A.6}T^{(2)}_l(r)&=&\int^\pi_0\vc{v}^*_{l+1}(\vc{r})\cdot\nabla\vc{v}_{l}(\vc{r})\cdot\vc{e}_z\sin\theta\;d\theta\nonumber\\
&=&\frac{2l+2}{2l+3}\big[-(l+1)(l+2)\frac{(f^*+g^*)g}{r}-(l+1)f^*g'+(l+2)g^*g'\big].
\end{eqnarray}
In the calculation of the matrix $\du{B}_B$ we need the angular integral
  \begin{eqnarray}
\label{A.7}V^{(1)}_l(r)&=&\int^\pi_0\vc{v}^*_l(\vc{r})\cdot(\nabla\vc{v}_{l+1}(\vc{r}))\cdot\vc{B}_1\sin\theta\;d\theta\nonumber\\
&=&\frac{2l+2}{(2l+1)(2l+3)}\big[-l(l^2+l-3)\frac{(f^*+g^*)f}{r}+3l(l+2)(l+3)\frac{(f^*+g^*)g}{r}\nonumber\\&-&l^2f^*f'+l(l+1)g^*f'+3l(l+2)f^*g'-3(l+1)(l+2)g^*g'\big],
\end{eqnarray}
as well as the integral
  \begin{eqnarray}
\label{A.8}V^{(2)}_l(r)&=&\int^\pi_0\vc{v}^*_{l+1}(\vc{r})\cdot(\nabla\vc{v}_{l}(\vc{r}))\cdot\vc{B}_1\sin\theta\;d\theta\nonumber\\
&=&\frac{2l+2}{(2l+1)(2l+3)}\big[-3l(l-1)(l+2)\frac{(f^*+g^*)f}{r}+(l+2)(l^2+3l-1)\frac{(f^*+g^*)g}{r}\nonumber\\&-&3l(l+1)f^*f'+3l(l+2)g^*f'+(l+1)(l+2)f^*g'-(l+2)^2g^*g'\big].
\end{eqnarray}
The above expressions can be derived by use of Legendre function identities.

\section{\label{B}}
In this Appendix we list the expressions for the matrix elements as functions of $s$ for low multipole orders. We consider the $5\times 5$ Hermitian matrices $\hat{\du{A}}_{13},\;\hat{\du{B}}_{S13}$ and $\hat{\du{B}}_{B13}$ corresponding to multipole coefficients $\mu_1,\kappa_2,\mu_2,\kappa_3,\mu_3$ in the representation of Sec. III. For brevity we denote the matrix elements as $A_{\alpha\beta},\;B_{S\alpha\beta},\;B_{B\alpha\beta}$ with $(\alpha,\beta)=1,...,5$.  We list only nonvanishing elements. The elements of the matrices $\hat{\du{A}}_{12},\;\hat{\du{B}}_{S12}$ and $\hat{\du{B}}_{B12}$ are found by deleting the fourth and fifth rows and columns.

The nonvanishing elements $A_{\alpha\beta}$ are given by
  \begin{eqnarray}
\label{B.1}A_{11}&=&3,\nonumber\\
A_{22}&=&\frac{15}{8s^8}\big[180+360(s+s^2)+240s^3+117s^4+42s^5+10s^6+2s^7\big],\nonumber\\
A_{23}&=&A^*_{32}=\frac{3}{s^4}\big[15+15(1+i)s+12is^2-2(1-i)s^3\big],\nonumber\\
A_{33}&=&6,\nonumber\\
A_{44}&=&\frac{21}{16s^{10}}\big[23625+47250(s+s^2)+31500s^3+15606s^4+6012s^5+1812s^6+408s^7+64s^8+8s^9\big],\nonumber\\
A_{45}&=&A_{54}^*=\frac{15-15i}{4s^5}\big[105+105(1+i)s+90is^2-20(1-i)s^3-4s^4\big],\nonumber\\
A_{55}&=&10.
\end{eqnarray}
The nonvanishing elements $B_{S\alpha\beta}$ are given by
  \begin{eqnarray}
\label{B.2}B_{S12}&=&B_{S21}^*=\frac{1}{2s^4}\big[-45i-45(1+i)s-42s^2-12(1-i)s^3+4is^4\big],\nonumber\\
B_{S13}&=&B_{S31}^*=-3i,\nonumber\\
B_{S24}&=&B_{S42}^*=\frac{9-9i}{8s^9}\big[1575+3150 s+(3150-90i)s^2+(2100-180i)s^3+(1032-180i)s^4\nonumber\\&+&(384-120i)s^5+(104-56i)s^6+16(1-i)s^7\big],\nonumber\\
B_{S25}&=&B_{S52}^*=\frac{3}{7s^4}\big[-105i+105(1-i)s+90s^2+20(1+i)s^3+4is^4\big],\nonumber\\
B_{S34}&=&B_{S43}^*=\frac{9-9i}{20s^5}\big[525+525(1-i)s-480is^2-130(1+i)s^3-44s^4-4(1-i)s^5\big],\nonumber\\
B_{S35}&=&B_{S53}^*=-6i.
\end{eqnarray}
The nonvanishing elements $B_{B\alpha\beta}$ are given by
  \begin{eqnarray}
\label{B.3}B_{B12}&=&B_{B21}^*=\frac{1}{2}\big[-i-(1+i)s+s^2-(1-i)s^3-2is^4F_-\big],\nonumber\\
B_{B24}&=&B_{B42}^*=\frac{1-i}{4s^5}\big[45+90s+(90-6i)s^2+(60-12i)s^3-(6+12i)s^4\nonumber\\&+&(4+24i)s^5-(4+48iF_2)s^6+8s^7-16s^8F_2\big],\nonumber\\
B_{B25}&=&B_{B52}^*=\frac{1}{168}\big[-18i+(18-18i)s+6s^2-(6+6i)s^3+9is^4\nonumber\\&+&(11-11i)s^5+(1-24F_+)s^6-(1+i)s^7+2is^8F_+\big],\nonumber\\
B_{B34}&=&B_{B43}^*=\frac{1-i}{480s}\big[450+450(1-i)s-222is^2+78(1+i)s^3\nonumber\\&-&81s^4+83(1-i)s^5-is^6+(1+i)s^7+168is^6F_--2s^8F_-\big],\nonumber\\
\end{eqnarray}
with the abbreviations
\begin{equation}
\label{B.4}F_+=F(s+is),\qquad F_-=F(s-is),\qquad F_2=F(2s)
\end{equation}
for values of the function $F(z)$ defined in Eq. (7.1).
The vanishing of the elements $B_{B13}$ and $B_{B35}$ follows from a general theorem on the potential flow contribution which we derived earlier \cite{16}.

The matrix elements have a quite complicated dependence on the scale number $s$. The limiting behavior of the maximum eigenvalues $\lambda_{12}(s)$ and $\lambda_{13}(s)$
for large $s$, shown in Figs. 1 and 2, is due to delicate cancellations.

\newpage

\newpage

\section*{Figure captions}

\subsection*{Fig. 1}
Plot of the maximum eigenvalue $\lambda_{12}(s)$ of a swimmer with three multipolar modes for $l=1,2$ as a function of the scale number $s$, defined in Eq. (3.13) (solid curve). The function has a maximum at $s_0=0.865$. We compare with the maximum value $1/\sqrt{2}$ for potential swimming with two modes for $l=1,2$ (dashed line).

\subsection*{Fig. 2}
Plot of the maximum eigenvalue $\lambda_{13}(s)$ of a swimmer with five multipolar modes for $l=1,2,3$ as a function of the scale number $s$ (solid curve). The function has a maximum at $s_0=0.962$. We compare with the maximum value $\sqrt{11/10}$ for potential swimming with three modes for $l=1,2,3$ (dashed line).

\subsection*{Fig. 3}
Plot of the function $\rho_{12}(s)$, as defined in Eq. (8.2), as a function of the scale number $s$.

\subsection*{Fig. 4}
Plot of the deformed sphere with surface displacement $\vc{\xi}(\vc{s})=\varepsilon\mathrm{Re}[-ia(\mu^I_1\vc{u}_1(\vc{s})+\kappa^I_2\vc{v}_2(\vc{s})+\mu^I_2\vc{u}_2(\vc{s}))\exp(-i\omega t)]$ with $\vc{v}_2(\vc{s})$ given by Eq. (7.2) of Ref. 18, with coefficients $(\mu^I_1,\kappa^I_2,\mu^I_2)=\vc{\xi}^0_0=(1,-4i\sqrt{2}/3,11i/(5\sqrt{2})$, amplitude factor $\varepsilon=0.15$, at times $t=0,\;\pi/(6\omega),\;\pi/(3\omega),\;\pi/(2\omega)$.

\newpage
\setlength{\unitlength}{1cm}
\begin{figure}
 \includegraphics{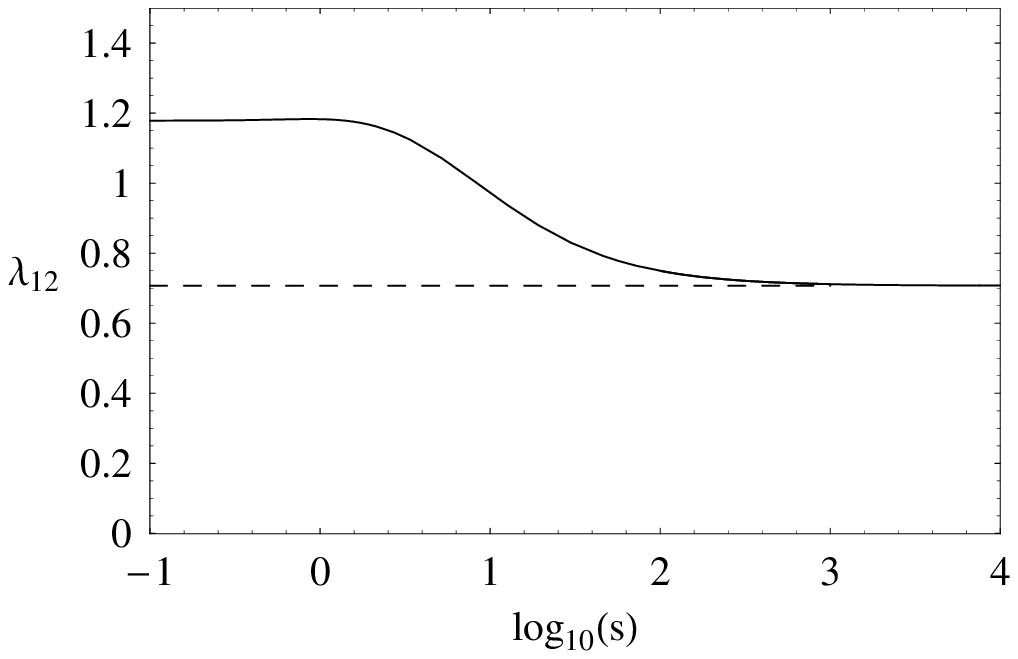}
   \put(-9.1,3.1){}
\put(-1.2,-.2){}
  \caption{}
\end{figure}
\newpage
\clearpage
\newpage
\setlength{\unitlength}{1cm}
\begin{figure}
 \includegraphics{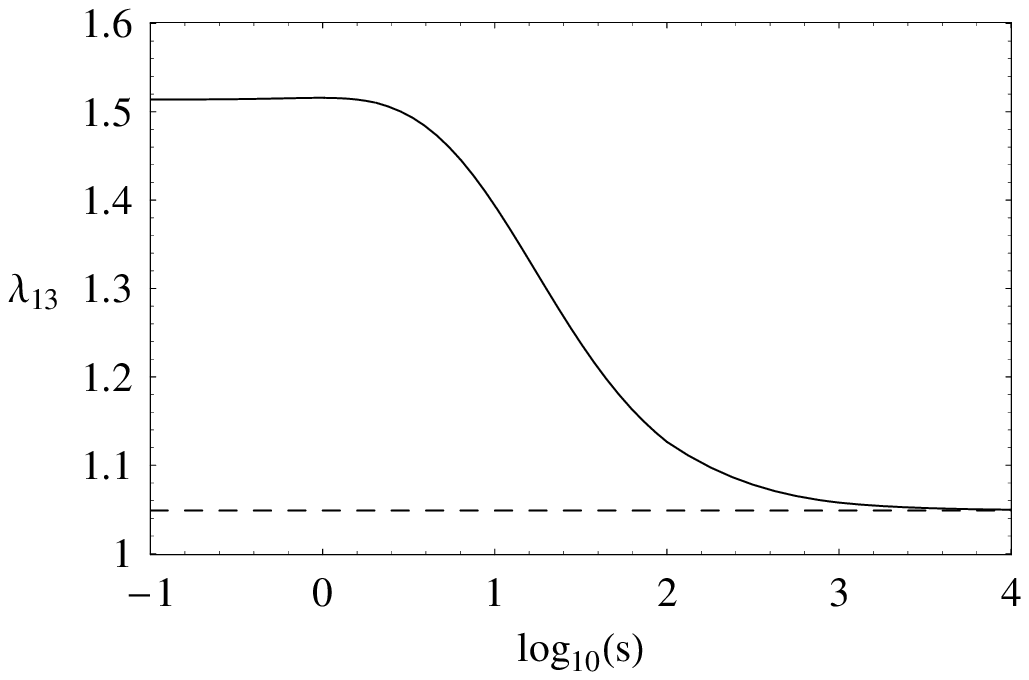}
   \put(-9.1,3.1){}
\put(-1.2,-.2){}
  \caption{}
\end{figure}
\newpage
\clearpage
\newpage
\setlength{\unitlength}{1cm}
\begin{figure}
 \includegraphics{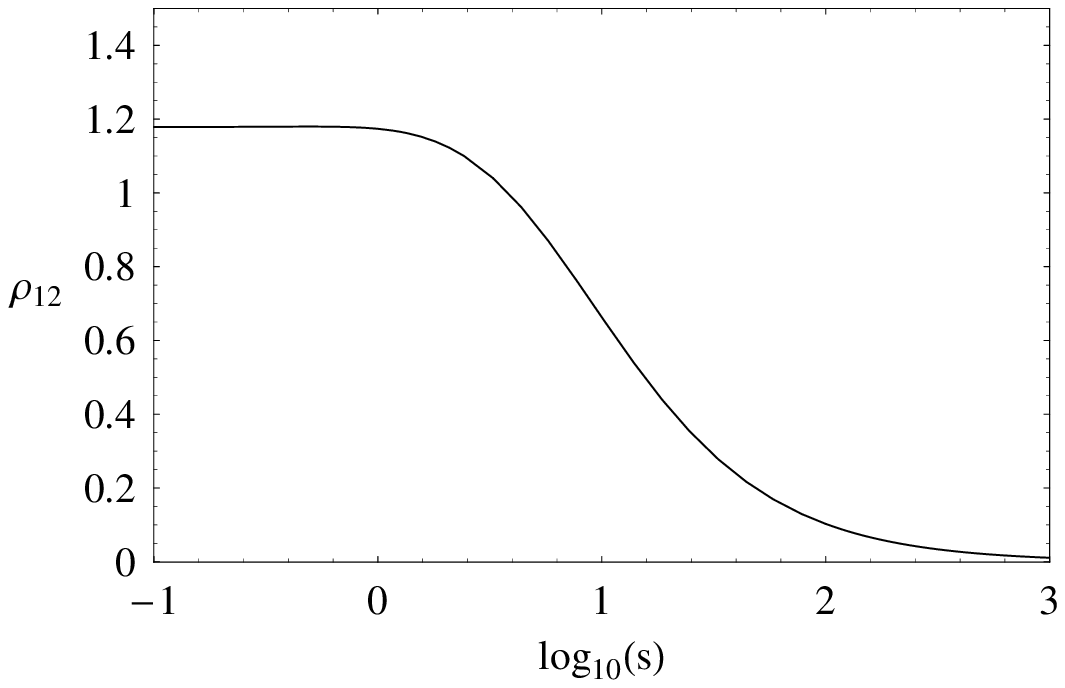}
   \put(-9.1,3.1){}
\put(-1.2,-.2){}
  \caption{}
\end{figure}
\newpage
\clearpage
\newpage
\setlength{\unitlength}{1cm}
\begin{figure}
 \includegraphics{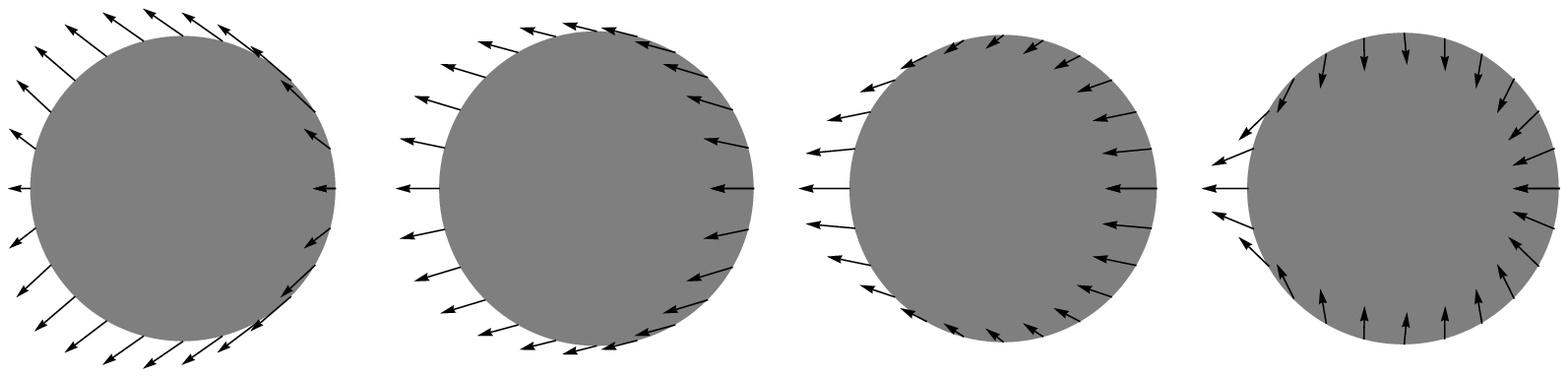}
   \put(-9.1,3.1){}
\put(-1.2,-.2){}
  \caption{}
\end{figure}
\end{document}